# Analysis by the Two-fluids model of the Dynamical Behavior of a Viscoelastic Fluid Probed by Dynamic Light Scattering.

*Eric MICHEL, Grégoire PORTE, Luca CIPELLETTI, Jacqueline APPELL*[*]

Groupe de Dynamique des Phases condensees, UMR5581 CNRS-Université Montpellier II
C.C.26, F-34095 Montpellier Cedex 05, France

*To whom correspondence should be adressed : appell@gdpc.univ-montp2.fr

ABSTRACT :The dynamic properties of a model transient network have been studied by dynamic light scattering. The network is formed by microemulsions droplets linked by telechelic polymers (modified hydrophilic polymers with two grafted hydrophobic stickers). We compare the properties of two networks similar but for the residence time of the hydrophobic stickers in the droplets. The results are interpreted according to the so-called two-fluids model, which was initially developed for semi-dilute polymer solutions[1-5] and which we extend here to any Maxwellian viscoelastic medium characterized by its elastic modulus and terminal time as measured by rheology. This model is found to describe consistently and quantitatively the experimental observations.

## INTRODUCTION

A number of complex fluids have viscoelastic properties: under a given stress they behave, at short times, as an elastic solid characterized by an instantaneous elastic modulus $G_0$ then at longer time they begin to flow and eventually behave as a liquid of viscosity η ($\eta = \tau_R * G_0$ where $\tau_R$ is the rheological terminal time). These complex fluids are generally described as some kind of transient network embedded in a solvent. Such a viscoelastic medium scatters light and this is essentially due to concentration fluctuations. The relaxation of these concentration fluctuations is more complex than the simple diffusion of usual Brownian dispersions and can be studied by dynamic light scattering (DLS). The dynamics probed by DLS reflects, in some way, the viscoelastic character of the complex fluid. At short times the relaxation is dominated by mutual diffusion. At long times a second relaxation mode is found for which diffusion is not the rate limiting step. Indeed the characteristic time of this second mode is of the order of the rheological terminal time. Brochard and De Gennes[1,2] and later Wang[3,4] and Genz[6] described theoretically the interplay between mutual diffusion and viscoelastic properties for semidilute polymer solutions for which the rheological properties have been described by Doi and Edwards[7]. The predictions of the theoretical model known as the two-fluids model have been experimentally checked on different polymer solutions by Adam and Delsanti[5], Brown et al[8] and Nicolai et al[9]. The DLS spectra for other viscoelastic materials such as surfactant giant micelles[10-13], thermoreversible aqueous copolymer gel[14] and solutions of associating random block copolymer[15,16] or telechelic ionomers[17] have also been studied.

A transient network of telechelic polymers (water-soluble polymers with hydrophobic endblocks or oil-soluble polymers with hydrophilic endblocks) also forms a viscoelastic medium. In binary solutions, the viscoelastic properties are controlled by the density of the network formed by the hydrophobic clusters of endblocks linked by the hydrosoluble polymers. A number of studies [18-24] have been devoted to these solutions of great practical interest. Alami et al[25] and Chassenieux et al[21] observe, in DLS, two and sometimes three relaxation times (even in binary systems). In order to better control the properties of these transient networks and to be closer to the conditions of their practical applications, Gradsielski et al[26] and Bagger-Joergensen et al[27] have initiated the studies of ternary systems: an oil-in-water microemulsion to which a telechelic polymer is added (inverse systems consisting of water-in-oil microemulsions linked by an hydrophobic polymer grafted at both ends with hydrophilic endblocks have also been studied[28-30]). The hydrophobic endblocks(stickers) stick in the hydrophobic core of the microemulsion droplets while the hydrophilic chains link or decorate the droplets: the resulting transient network is the origin of the viscoelastic properties of these fluids (see fig. 1). Both direct[31] and in inverse[28-30] systems have been studied by DLS. Two or three relaxation modes are observed depending on the particular system and the origin of these modes are still debated. In some cases all three relaxations are dominated by diffusion[28-30] while in other cases [16,31] the fast and the slow modes are dominated by diffusion while the intermediate one is not.

These ternary systems represent an ideal model system for studying the dynamics of viscoelastic networks. Indeed, contrary to the binary solutions recalled above, they allow one to control separately the average distance between the nodes of the network and the average degree of connections between these nodes. In fact the surfactant and oil concentration monitors the number density of the microemulsion droplets i.e. the nodes (the size of which is constant and essentially controlled by the self-assembling properties of the surfactant component) while the number of telechelic polymers per droplet determines the connectivity of the network. Moreover the contribution of the droplets to the scattering of light or neutron heavily dominates that of the polymer, making thus the analysis of scattering data easier and less ambiguous than in other systems. Finally, we point out that the time scale of the viscoelastic relaxation is controlled by the average residence time of the stickers in the droplets and can thus be readily varied over orders of magnitude by using telechelic polymers with aliphatic chains (the stickers) of different lengths, as described in the following.

Our group has been working on these ternary systems for several years. Structural characterization by small angle neutron scattering (SANS) has shown that the size and shape of the droplets are not affected by the addition of the telechelic polymer. The evolution of the structure factor upon addition of the polymer and, in parallel, the phase behavior of the systems clearly indicate that the polymers tend to bridge neighboring droplets[32,33]. Consistently with this structural picture of droplets reversibly linked to one another (see fig. 1), the samples exhibit strong viscoelastic behavior provided that the polymer amount is large enough. Their dynamical properties have been studied by rheology, DLS [34-36], and Fluorescence Recovery after Photobleaching (FRAP)[37].

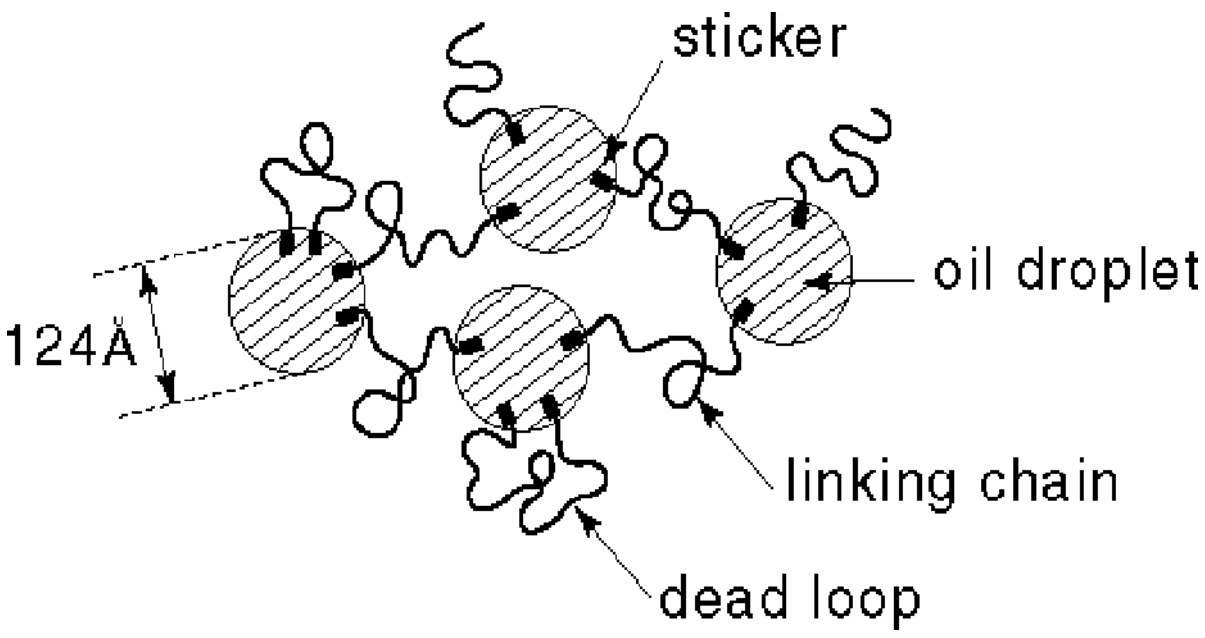

Figure 1 Representation of the viscoelastic transient network. The microemulsion droplets (decane droplets stabilized by a surfactant film in water, radius(~84>) are linked by the telechelic polymers, PEO, endgrafted by aliphatic chains. The dead loops (telechelic polymers decorating one single droplet) do not contribute to the viscoelastic properties of the network.

We argued in ref 34, that one needs at least three distinct parameters to describe the local state of the system, the volume fraction of droplet $\Phi$, the number of polymers per droplet, r/2, and the ratio of links to loops (see fig.1). DLS probes the fluctuations of $\Phi$, but we stress that $\Phi$ is coupled to the two other parameters. Accordingly three relaxation modes are to be expected. Moreover, since $\Phi$ and r are conserved variables (the total amount of droplets and of polymers is constant) while the ratio of links to loops is not conserved as it can change locally without transport of matter. We thus expect two modes to be dominated by diffusion ( characteristic time $\tau \sim q^{-2}$ ) and the remaining to be q-independent (q is the scattering vector). Indeed, among the systems we have studied, one (with stickers made of $C_{12}$ aliphatic chains: the C12 - system) displays three modes[37] , while in the second one ( the same system but with stickers made of $C_{18}$ aliphatic chains: the C18 - system) we had clear evidence of only two modes, the fastest one being diffusive while the second one is almost q independent[34]. Both systems have the same phase behavior and the same structure, as evidenced by SANS, the only difference lies in the average residence time of the stickers which is three orders of magnitude larger for the C18 - system than for the C12-system. The third relaxation mode observed in the C12-system contributes only for 10% of the total scattering. It is diffusive and its characteristic time is comparable to that measured by FRAP (fluorescence recovery after photobleaching) which probes the self-diffusion of the droplets. We argued in ref 37 that it is due to a relaxation via the interdiffusion of the droplets and of the polymers in the viscoelastic medium in good agreement with the conclusion reached by Schwab and Stühn[31] for their results on a similar system. A third relaxation mode is not observed clearly in the C18-system. It can be expected to be on a time scale difficult to observe experimentally, it probably accounts for a small part of the scattering as in the C12 -system and can possibly be mixed up with the second mode. On the other hand the much larger residence time of the stickers in the C18-system leads to viscoelastic properties ($G_0$ , $\tau_R$ ) readily measured by rheology [34,36] while they are inaccessible for the C12-system.

In this paper we extend the two-fluids model to any viscoelastic medium characterized by $G_O$ and $\tau_R$ . We then compare the predictions of the model to the experimental light scattering spectra measured for samples of the C12 and the C18 systems, which only differ by the time scale of the viscoelastic relaxation as mentioned above. We focuse here on the two relaxations modes observed in both samples. The fastest mode is diffusive and relaxes at a rate of the same order of magnitude in both systems; qualitatively it can be pictured as being dominated by the diffusion of the droplets at constant local links to loops ratio (diffusion at quenched topology). The slowest mode relaxes 1400 times faster in the C12-system, in good agreement with the estimated ratio of the residence time for C12 versus C18 stickers (see below). This result strongly suggest that this mode is coupled to the relaxation of the local links to loops ratio and is a strong indication that the second mode in the C12- and the C18-systems is the same "viscoelastic" mode (diffusion at annealed topology). Overall, we find a good quantitative agreement between the predictions of the two-fluids model and the experimental results for the two systems.

## THE TWO-FLUIDS MODEL

In permanent gels the scattered intensity comprises two contributions, a static contribution arising from frozen heterogeneities of the gel and a dynamic contribution arising from therma l fluctuations[40,41]. A transient gel, such as the system studied here, is ergodic therefore the static contribution to the light scattering is zero. Semi-dilute polymer solutions are other examples of transient gels, viscoelastic and ergodic as our system.

The model developed in[1,2,5] for semi-dilute polymer solutions can be reformulated more generally for a Maxwellian viscoelastic fluid characterized in rheology by an instantaneous elastic modulus $G_0$ and a terminal time $\tau_R$. In this section, we first derive a generalized diffusion equation for such a medium, and then calculate the time autocorrelation function of the concentration fluctuations described by this equation.

a/ Relaxations of the fluctuations of concentration in the framework of the two-fluids model.

The viscoelastic fluid is pictured as a suspension of particles with a mean number density $n_o$ embedded in a viscoelastic medium. The viscoelastic medium is characterized by an instantaneous elastic modulus $G_0$ and a terminal time $\tau_R$ .The local number density fluctuates and can be written at point **r** and time t as: $n(\mathbf{r},t) = n_o + \delta n(\mathbf{r},t)$. The free energy density of the particles writes: $g(n) = g(n_o) + \frac{1}{2}\frac{\partial^2 g}{\partial n^2}\delta n^2$ and the particles move in order to minimize g(n). A particle in **r** at time t will be in **r'** at time t': $\mathbf{r}' = \mathbf{r} + \mathbf{u}(\mathbf{r})$ , $\dot{\mathbf{u}} = \frac{\delta \mathbf{u}}{\delta t}$ .

The conservation equation writes $\frac{\partial n}{\partial t} = -\nabla\left(n\dot{\mathbf{u}}(\mathbf{r},t)\right) \Rightarrow \frac{\partial \delta n(\mathbf{r},t)}{\partial t} = -n_o \nabla \dot{\mathbf{u}}(\mathbf{r},t)$

The number density fluctuations give rise to an osmotic force $F_0$ which sets the particles to move in order to sweep out the fluctuations. In a purely viscous fluid, a viscous force $F_v$ resists this motion, while in a viscoelastic fluid a transient viscoelastic force $F_e$ also come into play. These forces can be written:

$$\mathbf{F}_o = -n_o \frac{\partial^2 g}{\partial n^2} \nabla \delta n(\mathbf{r},t)$$

$\mathbf{F}_v = -n_o f \dot{\mathbf{u}}$ $f$ is the frictional coefficient. In the limit of high dilution, $f = 6\pi\eta R$ for a particle of hydrodynamic radius R in a fluid of viscosity η (Stokes' law)

$\mathbf{F}_e = \int_{-\infty}^{t} \left[ \left( K(t'-t) + \frac{1}{3} G(t'-t) \right) \nabla(\nabla.\dot{\mathbf{u}}(t')) + G(t'-t)\Delta \dot{\mathbf{u}}(t') \right] dt'$ where K and G are respectively the isostatic compressibility and the shear modulus [38,39].

Writing Newton's second law for the movement, neglecting the inertial term, taking into account the conservation of particles and taking the divergence of each term we obtain:

$$-n_o \frac{\partial^2 g}{\partial n^2} \Delta \delta n(\mathbf{r},t) + 6\pi\eta R \delta\dot{n}(\mathbf{r},t) - \frac{1}{n_o} \int_{-\infty}^{t} \left( K(t'-t) + \frac{4}{3} G(t'-t) \right) \Delta \delta\dot{n}(\mathbf{r},t') dt' = 0$$

The Fourier transform of the above equation is the diffusion equation and writes:

$$\delta\dot{n}(q,t) + D_{coll} q^2 \delta n(q,t) + D_{el} q^2 \int_0^t \exp\left(-\frac{t'}{\tau_R}\right) \delta\dot{n}(q,t-t') dt' = 0 \qquad (1)$$

In this diffusion equation, the two first terms correspond to the classical diffusion equation for brawnian motion in a viscous solvent, the third terms accounts for the additional transient elastic resistance of the network.

$D_{coll} = n_o \frac{\partial^2 g}{\partial n^2} \Big/ f$ is the usual collective translational diffusion coefficient of the particles and $D_{el} = 2G_o / n_o f$ is the viscoelastic diffusion coefficient, assuming $K_0 + \frac{4}{3} G_0 = 2G_0$ following the arguments developed by Tanaka, Hocker and Benedek[42] in their pioneering work on DLS by a polyacrylamide gel.

b/ <u>The autocorrelation function of the number density fluctuations</u> (that can be measured by DLS).

Solving the equation of diffusion, eq 1, by standard mathematics, we can derive the Fourier transform of the number density fluctuations $\delta n(q,t)$ and their autocorrelation function $g_1(t)$. It is found to be the sum of two simple exponential relaxations with well defined amplitudes $A_{f/s}$ and relaxation times $\tau_{f/s}$ where f and s indicate the fast and slow mode respectively. The result writes:

$$g_1(t) = \frac{\langle \delta n(q,t) \delta n(q,0) \rangle}{\left\langle |\delta n(q,0)|^2 \right\rangle} = \left[ A_f \exp(-\tau_f^{-1} t) + A_s \exp(-\tau_s^{-1} t) \right] \qquad (2)$$

With the amplitude and characteristic time of each mode given by

$$\tau_f^{-1} = \frac{1}{2}\left(\tau_R^{-1} + D_{eff} q^2\right)\left[1 + \sqrt{1 - \frac{4 D_{coll} q^2 \tau_R^{-1}}{(\tau_R^{-1} + D_{eff} q^2)^2}}\right], \quad A_f = -\frac{\left[D_{el} q^2 - \tau_f^{-1} + \tau_R^{-1}\right]}{\tau_f^{-1} - \tau_s^{-1}} \qquad (3a)$$

$$\tau_s^{-1} = \frac{1}{2}\left(\tau_R^{-1} + D_{eff} q^2\right)\left[1 - \sqrt{1 - \frac{4 D_{coll} q^2 \tau_R^{-1}}{(\tau_R^{-1} + D_{eff} q^2)^2}}\right], \quad A_s = \frac{\left[D_{el} q^2 - \tau_s^{-1} + \tau_R^{-1}\right]}{\tau_f^{-1} - \tau_s^{-1}} \qquad (3b)$$

where $D_{eff} = D_{coll} + D_{el}$ and $A_f + A_s = 1$

An hydrodynamic regime is found when the rheological terminal time is small or more precisely when $\tau_R << \left[D_{eff} q^2\right]^{-1}$. In this limit we find from Eq (2), to first order in $\tau_R^{-1}$, that $\tau_f \sim \tau_R$, $A_f \sim 0$ and $\tau_s \sim \left[D_{coll} q^2\right]$, $A_s \sim 1$. In this regime the fluctuations of concentration relax essentially by the diffusion of the droplets as if the telechelic polymer was not present. In the opposite limit, when the rheological terminal time is large ( $\tau_R >> \left[D_{eff} q^2\right]^{-1}$ ) Eqs (3a) and (3b) expanded to first order in $\tau_R^{-1}$ give:

$$\tau_f^{-1} = D_{eff} q^2 + \tau_R^{-1} \frac{D_{el}}{D_{eff}} \quad , \qquad A_f = \frac{D_{coll}}{D_{eff}} \qquad (4a)$$

$$\tau_s = \frac{1}{D_{coll} q^2} + \tau_R \frac{D_{eff}}{D_{coll}} \quad , \qquad A_s = \frac{D_{el}}{D_{eff}} \qquad (4b)$$

This is the "gel" regime characterized by two relaxation modes. One is fast, diffusive with a constant contribution proportional to $\tau_R$ , while the other is slow, proportional to $\tau_R$ and with a diffusive correction. In the intermediate regime $\tau_R \sim \left[D_{eff} q^2\right]^{-1}$, Eqs 3a and 3b have to be used to describe the two relaxations modes.

In the two-fluid model we assumed monodisperse droplets and a maxwellian relaxation therefore the model predicts, as it should, an overall relaxation which is the sum of two single exponentials.

For the two samples studied here, we expect $\tau_R$ 's to differ by more than three orders of magnitude while the diffusion coefficients are very close if not equal. Thus the two samples offer a unique opportunity to test the model. We will show in the following that the experimental results for the C18-system with the longest $\tau_R$ are well described by Eqs(4) □□□□□ □□□□□C12- system the exact form Eqs(3) yields a good agreement with the DLS data.

## EXPERIMENTAL SECTION

Preparation of samples

The samples are prepared by weight in triply distilled water. Decane from Fluka and the non-ionic surfactants TX100 and TX35 from Sigma Chemicals are used as received. The polymer is poly (ethylene-oxide), it has been hydrophobically modified and purified in the laboratory using the method described in [43,44]. The molecular weight of the starting products is determined by size-exclusion chromatography.

The hydrophobically modified poly(ethylene-oxide) contains an isocyanate group between the alkyl chain and the ethylene-oxide chain. We assume this isocyanate group to belong to the hydrophilic part of the copolymer. Two telechelic polymers have been prepared: poly (ethylene-oxide) PEO-2C12 with a $C_{12}H_{25}$ aliphatic chain grafted at each extremity and PEO-2C18 with a $C_{18}H_{37}$ aliphatic chain grafted at each extremity. After modification, the degree of substitution of the hydroxyl groups was determined by NMR using the method described in [45] .The degree of substitution is found to be equal or larger than 98%.

The microemulsion is a thermodynamically stable dispersion, in water, of oil droplets stabilized by a surfactant film[46]. We can adjust the size of the drops by varying the composition of the surfactant film (here the weight ratio of TX35 to TX100 is 0.5) which defines the spontaneous radius of curvature of the surfactant film and by choosing the weight ratio of decane to surfactant ( here 0.7) in order to be close to but slightly below the emulsification failure limit. Under such conditions it is well established that the droplets are spheres of a well-defined radius[47] . Indeed, we showed previously that the microemulsion droplets are spherical with a narrow distribution of size with a mean radius of 84±2Å and a standard deviation of 15 Å (for more details see ref 33) . An appropriate amount of PEO-2C12 or PEO-2C18 is then added to the microemulsion.

The samples are characterized by the volume fraction Φ of the microemulsion droplets, and by the number r of $C_{12}$ or $C_{18}$ chains per droplet. All the parameters to calculate Φ and r from the sample composition are summarized in Table 1.

Table 1 Molar Mass and density of the components of the samples

| Component (abbreviated in the text) | Molar Mass (dalton) | | Density ($g/cm^3$) | |
|---|---|---|---|---|
| | | HC(a) | polar part | HC(a) |
| $H_2O$ | 18 | - | 1 | - |
| $[H_3C-(C-(CH_3)_2-CH_2-C-(CH_3)_2)\varphi](O-CH_2-CH_2)_3-OH$ (TX35) | 338 | 189 | 1.2 | 0.86 |
| $[H_3C-(CH_2)_8CH_3]$ (decane) | 142 | 142 | - | 0.75 |
| $[CH_3-(CH_2)_{11}]$-NH-CO-$(O-CH_2-CH_2)_{227}$-O-(CO)-NH-$[(CH_2)_{11}\ CH_3]$. (PEO-2C12) | ~10 400 | 338 | 1.2 | 0.81 |
| $[CH_3-(CH_2)_{17}]$-NH-CO-$(O-CH_2-CH_2)_{227}$-O-(CO)-NH-$[\ (CH_2)_{17}CH_3]$. (PEO-2C18) | ~10 600 | 506 | 1.2 | 0.81 |

(a) HC= Hydrophobic part of the molecule in brackets in the formula on the left

For the C18-system, the evolution of the viscoelastic properties with r has been described and discussed previously[34] . This system has been shown to undergo a percolation transition with increasing polymer concentration. A percolation transition could not be evidenced directly for the C12-system because of the extremely short residence time of the stickers (we estimate $\tau_R$ to be of the order of $10^{-4}$ s., see below). However the great similarity of the two systems led us to argue that the same transition indeed exists in the C12-system[33] . In this paper, we focus on samples with Φ= 17.1% and r= 13.2 for both systems, far above the percolation threshold ($r_p$=3.1) and thus where the transient network is densely connected. The volume fraction Φhas been chosen so that the mean distance between droplets (center to center ~ 240Å) allows for the PEO chain linking two drops to have on average their "natural length" (the estimated end-to-end distance is 90Å[32]).

The difference in the two samples lies in the nature of the stickers: aliphatic chains with 12 or 18 carbon atoms; we will refer to them as the C12- or C18-sample. In the two samples the residence time $t_O$ of the stickers will differ, it is related to the adhesion energy $W$ of the stickers by the usual Arrhenius law:$t_0=w_0^{-1}\exp(W / k_BT)$, where $w_0$ is the attempt frequency of the order of the inverse self-diffusion time of the sticker. For a linear and saturated aliphatic chain, $W$ has been evaluated from critical concentration measurements of series of surfactants to be of the order of 1.2 $k_BT$ per $CH_2$ [48]. The residence time of a 12-carbon sticker inside a droplet should be thus about $10^3$ times shorter than that of a 18-carbon sticker. Since the viscoelastic time depends on the residence time of the stickers, we expect the ratio of terminal times to be equal to the ratio of residence times roughly equal to 1400. A value in excellent agreement with the ratio of 1360, we obtained by measuring the low-shear viscosity for identical samples of the two systems[37].

Dynamic Light Scattering

The measurements are performed on a standard setup (AMTEC Goniometer with a BI9400 Brookhaven correlator), the light source is an argon ion laser (λ=514.5 nm). The homodyne intensity autocorrelation function is measured at different q values, ranging from 3 $10^6$ to 3 $10^7$ $m^{-1}$ ( $q = \frac{4\pi n}{\lambda} \sin\frac{\theta}{2}$ with n the refractive index of the solvent and θ the scattering angle ). If the scattered field obeys Gaussian statistics, the normalized autocorrelation function of the scattered intensity $g_2(q,t) = \frac{\langle I(q,t) I(q,0)\rangle}{|\langle I(q,0)\rangle|^2}$ can be expressed as a function of the first-order electric field correlation function through the Siegert relationship $g_2(q,t) - 1 = c\,|g_1^E(q,t)|^2$ with c (0<c<1) an experimental constant. In our samples, the scattering originates from the fluctuations of the droplet concentration and $g_1^E(q,t) \propto g_1(q,t)$ given by Eq (2) .

A typical normalized autocorrelation function for the C18-sample measured at q= 2.3$10^7$ $m^{-1}$ is shown in fig. 2, the line is an adjustment to the following expression:

$$g_2(t) - 1 = \left[\alpha_f \exp(-\tau_f^{-1} t) + \alpha_s \exp\left(-(\tau_s^{-1} t)^{0.82}\right)\right]^2 \qquad (5)$$

where the slow relaxation mode , dominated by the viscoelasticity of the sample is represented by a slightly stretched exponential to parallel the stretched exponential found to be the best fit in the rheological measurement (see ref 34 and the illustration given below)

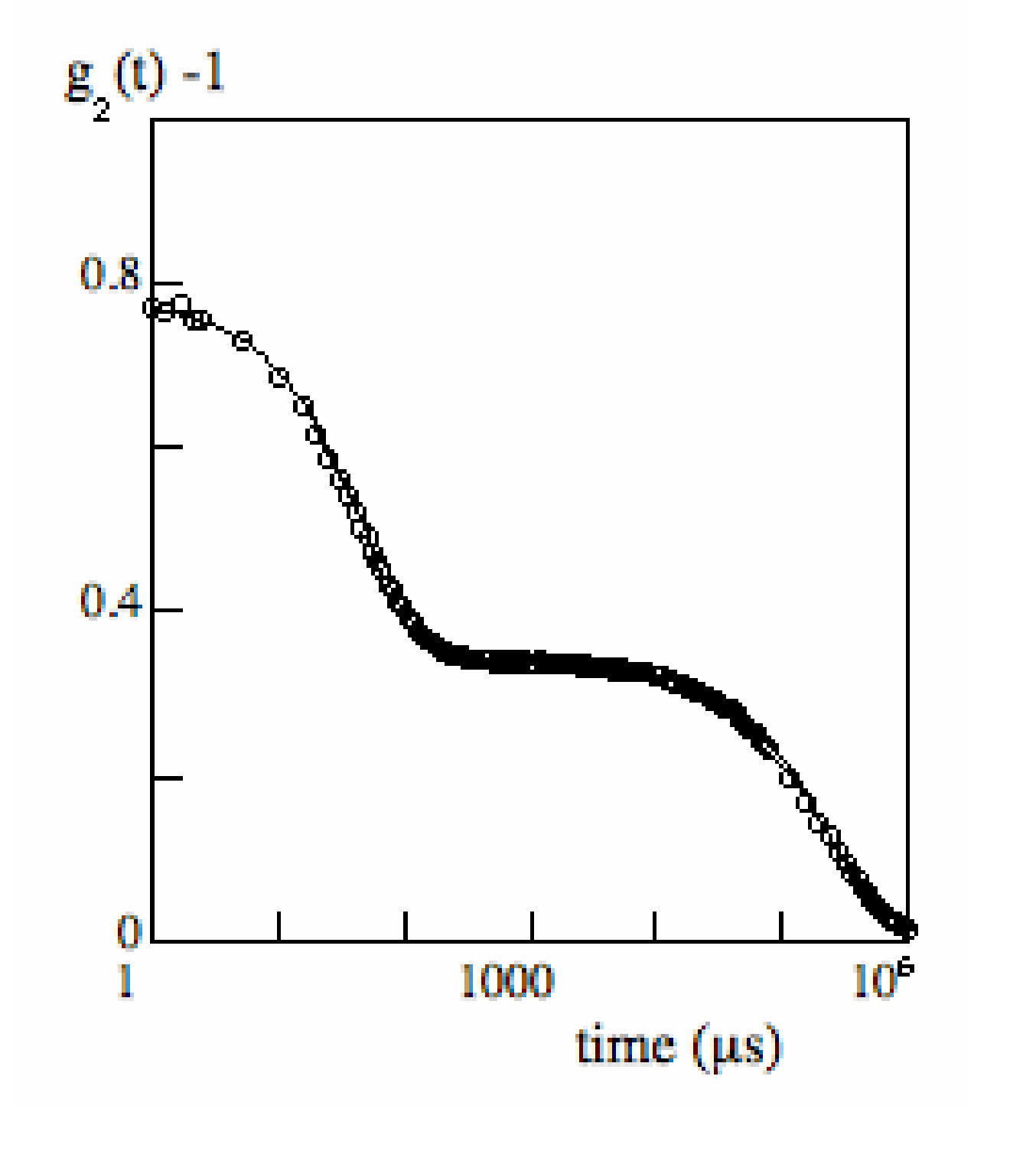

Figure 2 Normalized intensity autocorrelation function measured at q=2.3 $10^{7}$ $m^{-1}$ for the C18-sample. Open circles are the data points. The line is a fit to the data of Eq. (5), with $\alpha_f = 0.29$ , $\tau_f = 51\mu s$ , $\alpha_s = 0.59$ , $\tau_s = 0.56$ s

The autocorrelation function for the C18-sample can be compared to a typical normalized autocorrelation function for the C12-sample measured at the same angle and given in figure 3. Note that, as expected, the time scale extends to much smaller values than for the C18-sample. The best fit is obtained, as already mentionned, when three relaxation modes are taken into account:

$$g_2(t) - 1 = \left[ \alpha_f \exp\left(-\tau_f^{-1} t\right) + \alpha_s \exp\left(-(\tau_s^{-1} t)^{0.82}\right) + \alpha_p \exp\left(-\tau_p^{-1} t\right) \right]^2 \qquad (6)$$

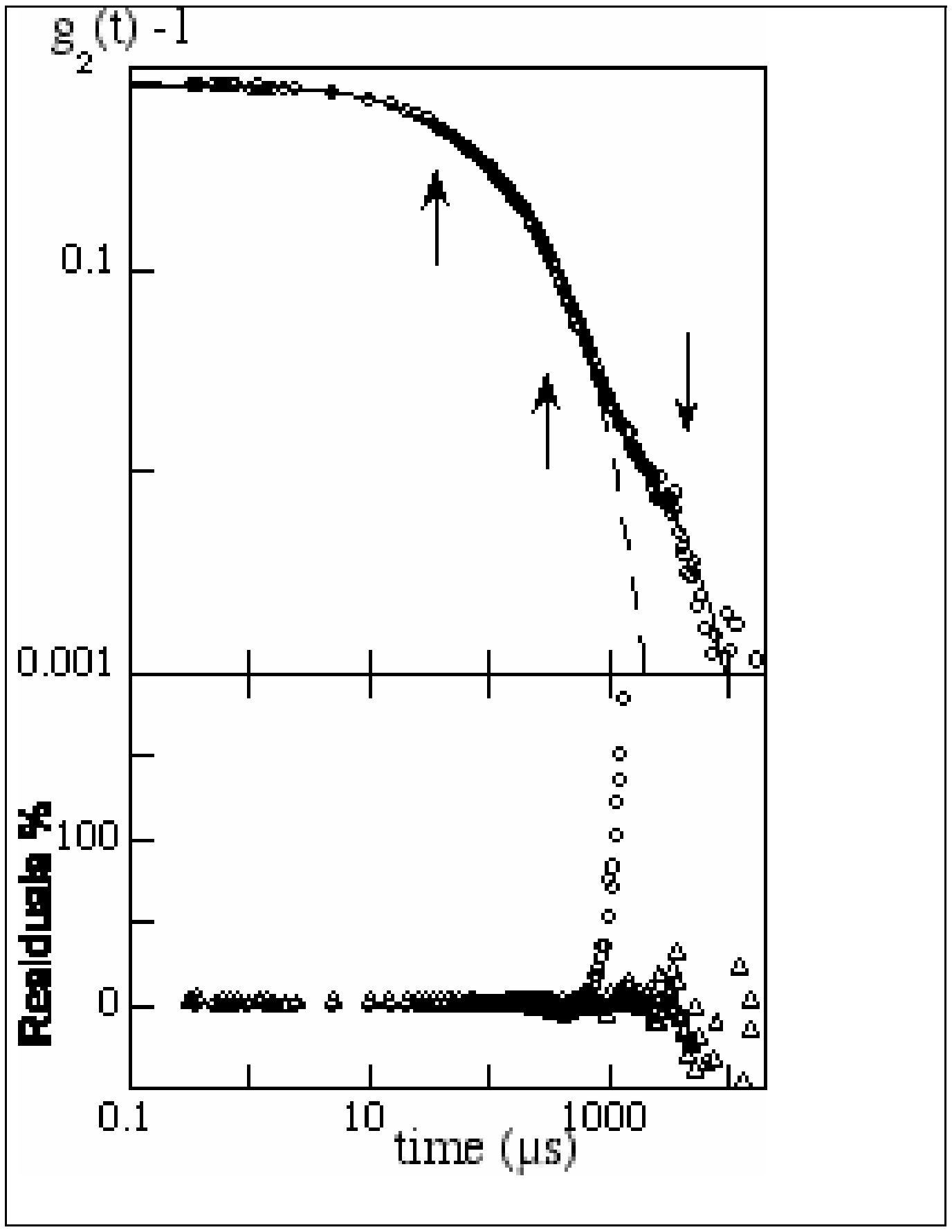

Figure 3 : Normalized intensity autocorrelation function measured at q=2.3 $10^7$ $m^{-1}$ for the C12-sample.The solid line shows the best fit to relation [6] for three relaxation modes. The arrows show the three relaxation times $\tau_f$ = 36μs, $\tau_s$ =300μs and $\tau_p$ = 6700μs,. The dashed line is the “ best ” fit to Eq. (5). The bottom graph shows the residuals: triangles (Eq. (6)) and circles (Eq. (5) )

Rheology : step strain experiments

To illustrate the viscoelastic behavior of the samples, the stress relaxation curve of the C18-sample is shown in fig. 4. It is obtained in a step strain experiment on a Rheometrics RFSII-strain-controlled rheometer. At time t=0 the sample is submitted to a sudden step shear strain (of amplitude ε small enough to be in the linear regime) and the shear stress response σ is recorded as a function of time. The decay of the time dependent modulus, $G(t) = \sigma(t)/\varepsilon,$ is best represented by a slightly stretched exponential[34]:

$$G(t) = G_0 \exp\left[-(t/\tau_R)^{0.82}\right] \qquad (7)$$

which corresponds to an almost Maxwellian behavior. We obtain $G_O$ = 1830 ± 100 Pa and $\tau_R$=~1□□ ± 0.005s.

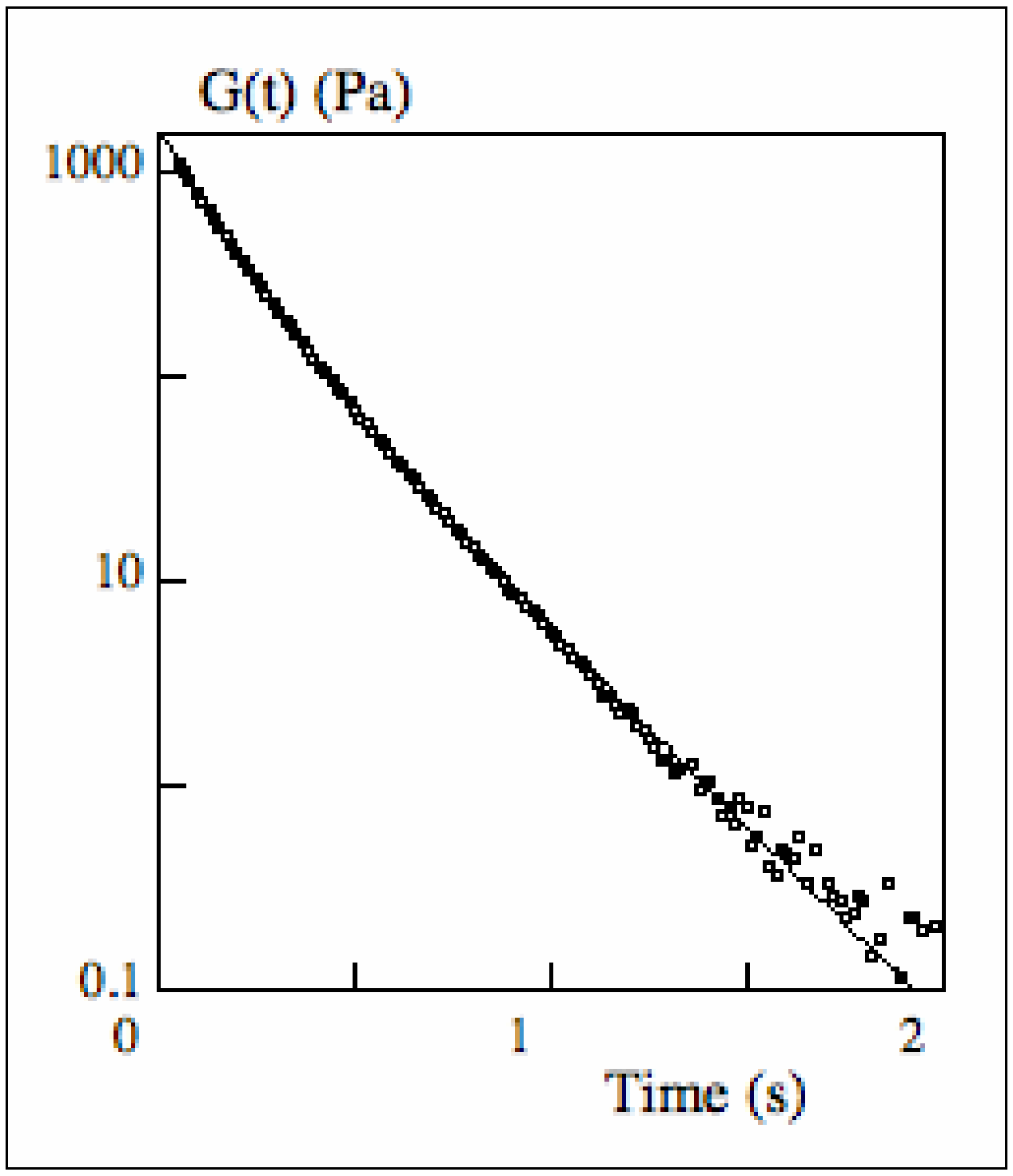

Figure 4 Stress relaxation curve for the C18-sample. The line is a fit to the data of a stretched exponential: Eq. (7), with $G_0$ = 1830± 100 Pa and $\tau_R$ = 0.125 ± 0.005 s.

**RESULTS AND DISCUSSION**

Measurements over the entire q-range allows us to follow the q-dependence of the characteristic times and of the relative amplitudes of the two modes. For the C18-sample $\tau_R \sim 0.125$s (as measured in rheology) is larger than the diffusion time $(D_{eff}q^2)^{-1}$ of the drops over the entire q-range; we thus probe the "gel" regime and expect Eqs (4) to hold. Indeed the fast relaxation mode is diffusive and its characteristic time is well described by relation (4a) with a negligible contribution of the constant term as shown in fig. 5 where $\tau_f^{-1}$ is plotted as a function of $q^2$, the slope of the straight line yields $D_{eff}$ = 3.3 $10^{-11}$ $m^2$ $s^{-1}$.

With this value of $D_{eff}$ and the value of $\tau_R$ the terminal time measured in the strain relaxation experiment (see fig.4) in the right handside of Eq (3b), the first term is found to be negligible compared to the second one and the characteristic time of the slow relaxation mode should not depend on q. However, as shown in fig. 6, a small dependency is found and will be briefly discussed below. Using the mean value $\tau_s$ =0.5s, we obtain from Eq (3b) an estimate of $D_{coll} \sim 0.9\ 10^{-11}$ $m^2$ $s^{-1}$ .

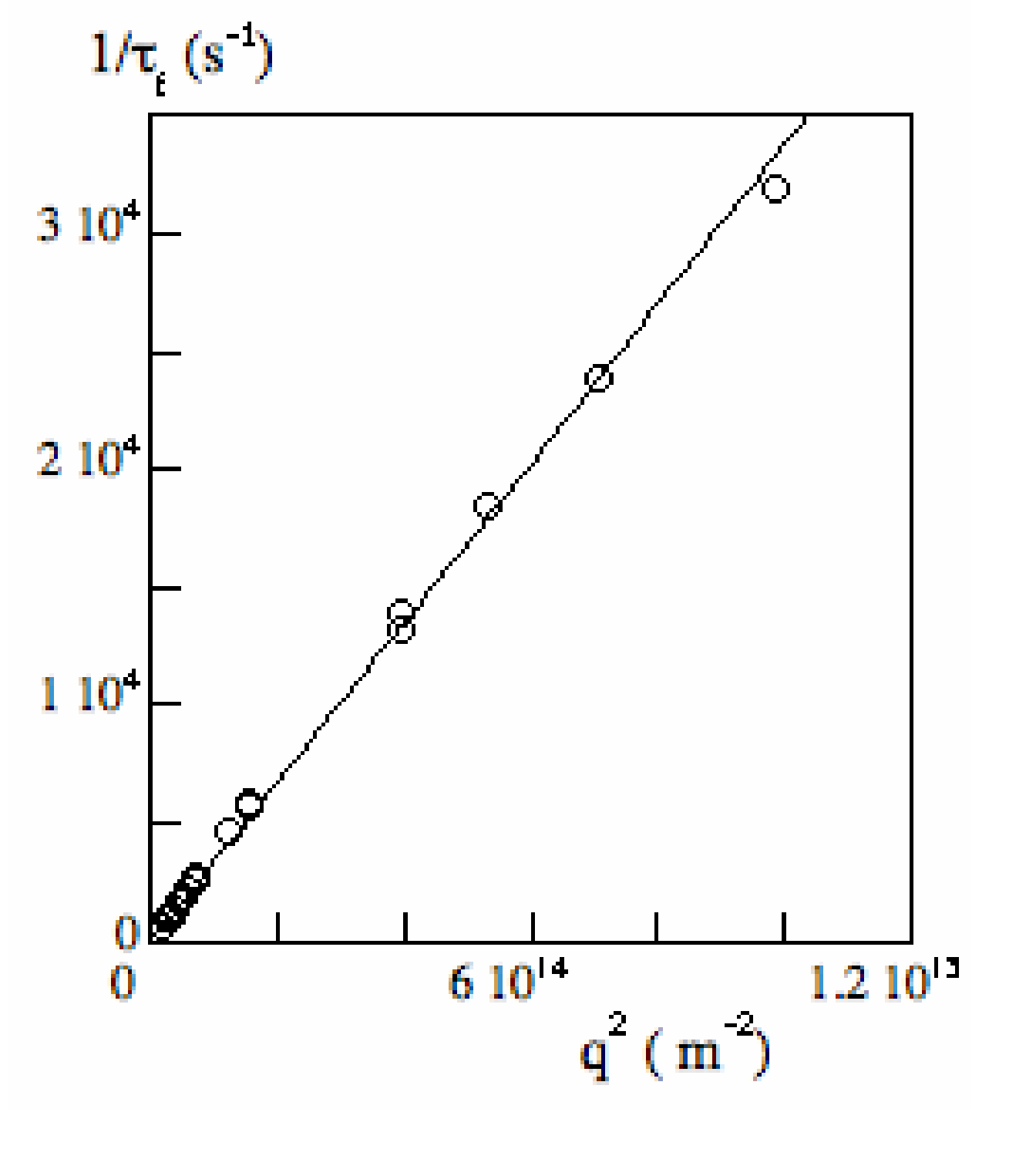

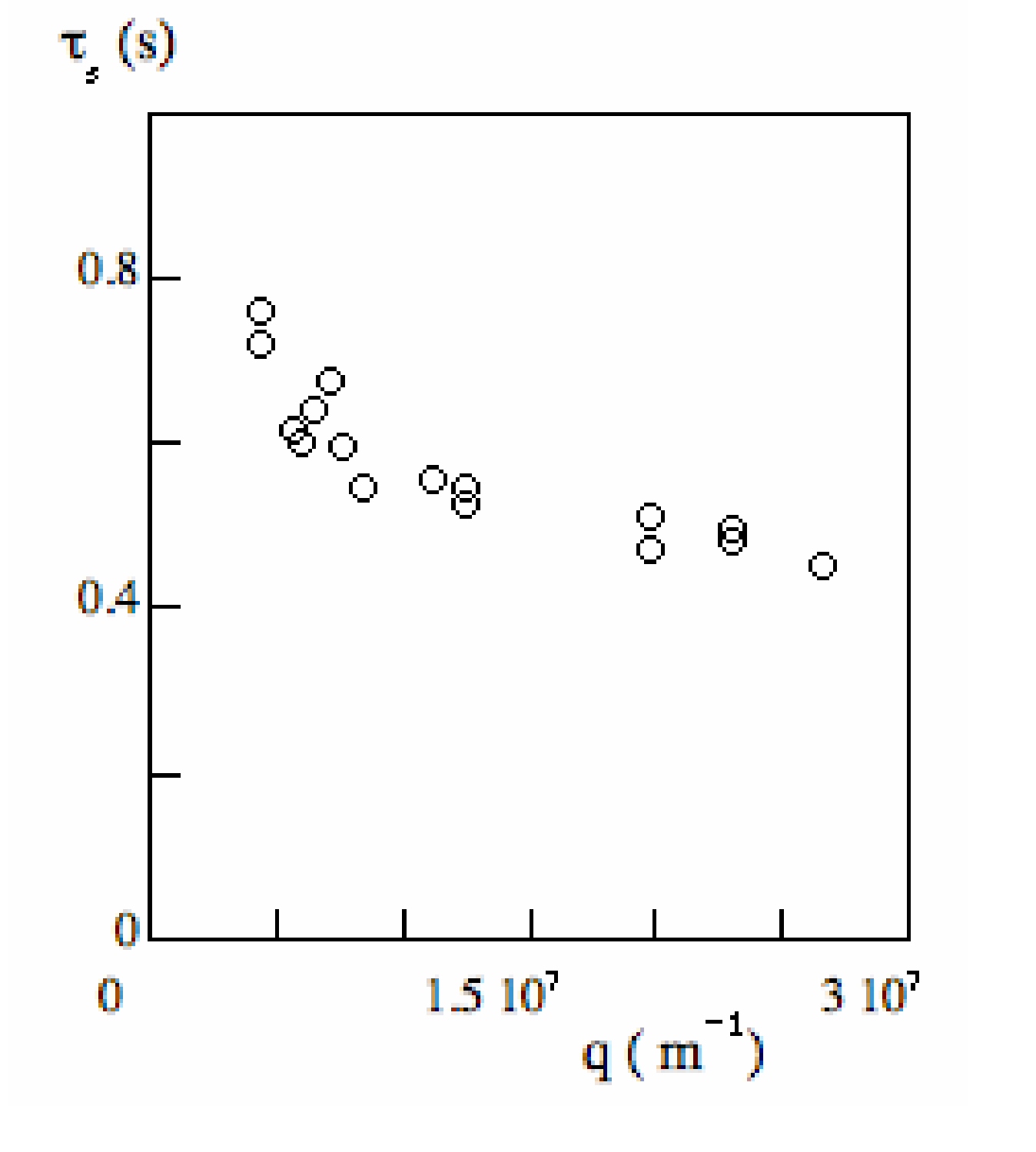

Figure 5 C18-sample: Illustration of the squared-q dependence of $\tau_f^{-1}$, the inverse relaxation time of the fast mode. The line is a linear fit to the data with a slope yielding $D_{eff} = 3.5\ 10^{-11}\ m^2\ s^{-1}$

Figure 6 C18-sample: Illustration of the weak dependence of $\tau_s$ the relaxation time of the slow mode on q

If we now examine the results for the C12-sample we find, as expected, that they cannot be interpreted in terms of the simplified expressions for the "gel" regime. A fit of Eqs (3) to the results for the characteristic times of both the fast and slow modes is shown in fig. 7 and yields the values of the parameters given in table 2. $\tau_R$ is found equal to 8.1 $10^{-5}$ s , 1500 times smaller than the value for the C18-sample ( $\tau_R$= 0.125s). This ratio is in excellent agreement with the evaluation of the ratio of the residence times given above for the C12 and C18 stickers. The values of $D_{eff}$ and $D_{coll}$ are almost equal (within experimental uncertainties) to those found for the C18-sample. We stressed above that the diffusion coefficients are expected to be equal for the two samples since they are identical as regard the concentration of droplets (Φ) and the connectivity of the network (i.e. the concentration of telechelic polymers r/2 and the ratio of links to loops). In the following discussion, we will adopt the mean values of these coefficients given in the last line of table 2.

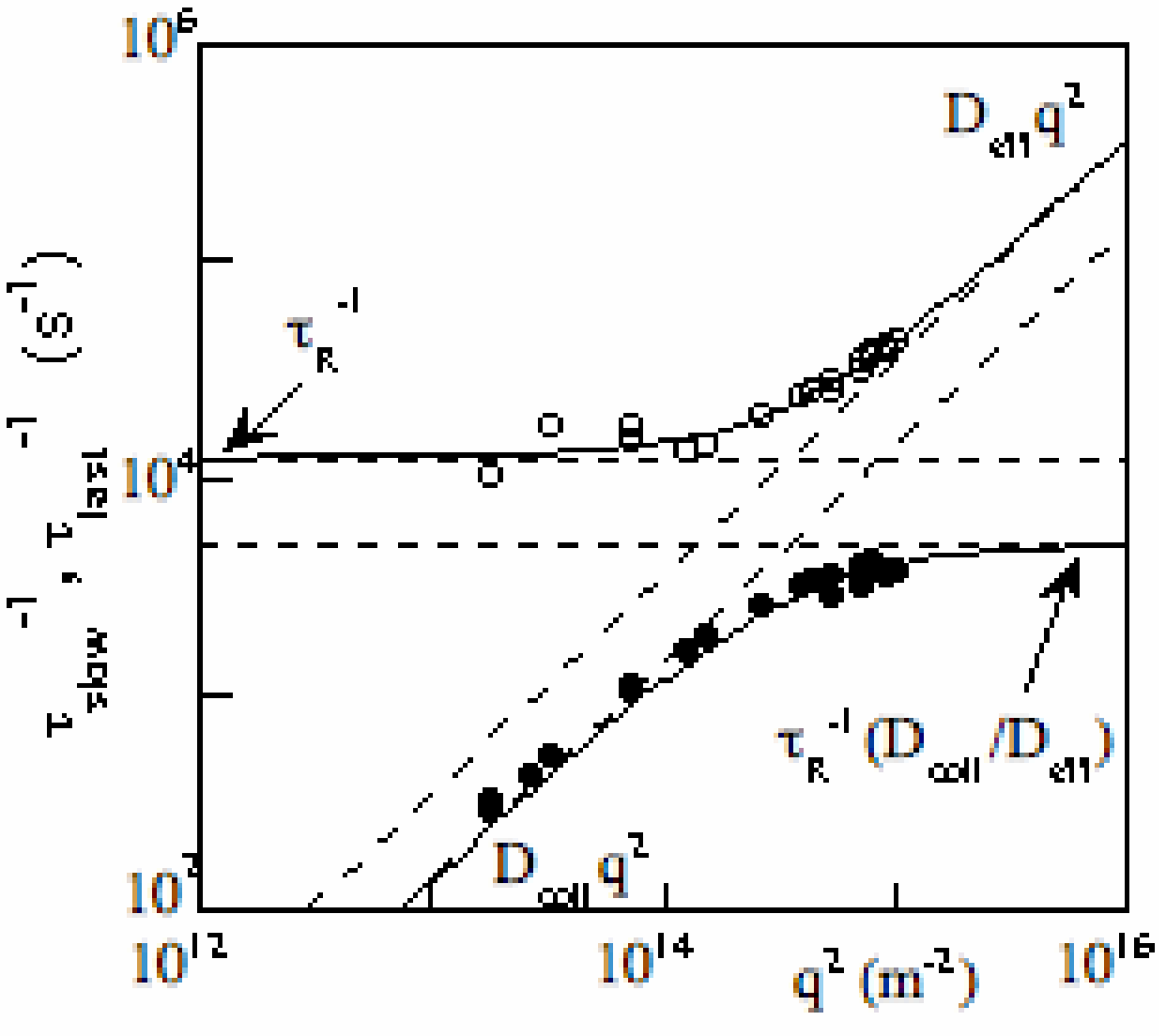

Figure 7 C12-sample: $q^2$ -dependence of the inverse of the fast (open circles) and slow (full circles) relaxation time . The solid lines are the simultaneous fit to the data of Eqs. (3). The fitting parameters are given in table 2. The dashed lines are the asymptotic behavior at small or large values of $q^2$.

Table 2 Values of the parameters deduced from the fit of Eqs. (2) and (3) to the relaxation times obtained by DLS. For the C18 sample, $\tau_R$ is measured by rheology.

| Parameter→<br>Experimental Sample↓ | $D_{eff}$ ($m^2$ $s^{-1}$ ) | $D_{coll}$ ( $m^2$ $s^{-1}$ ) | $\tau_R$ ( s ) |
|---|---|---|---|
| C12 | 3.5 ± 0.3 $10^{-11}$ | 1.4 ± 0.2 $10^{-11}$ | 8.1± 0.5 $10^{-5}$ |
| C18 | 3.3± 0.3 $10^{-11}$ | 0.9 ± 0.2 $10^{-11}$ | 0.125± 0.005 (Rheology) |
| mean value | 3.4 $10^{-11}$ | 1.2 $10^{-11}$ | - |

The coherence of the experimental results obtained on the two samples with their description by the two fluids-model are illustrated in figs. 8 and 9. In fig. 8, the characteristic times of the two modes in both samples are plotted as a function of $q^2$ in reduced units together with the theoretical curves calculated from Eqs (3) using $D_{eff}$ = 3.4 $10^{-11}$ $m^2$ $s^{-1}$ and $D_{coll}$ = 1.2 $10^{-11}$ $m^2$ $s^{-1}$. The agreement is excellent for the fast mode , it is reasonable for the slow mode. For the slow mode of the C18-sample we note here, as above, that the predicted q-dependence of the relaxation time is much less than that observed experimentally so that the discrepancy between theory and experiment is not simply due to the application of a more or less valid approximation to Eq (3b).

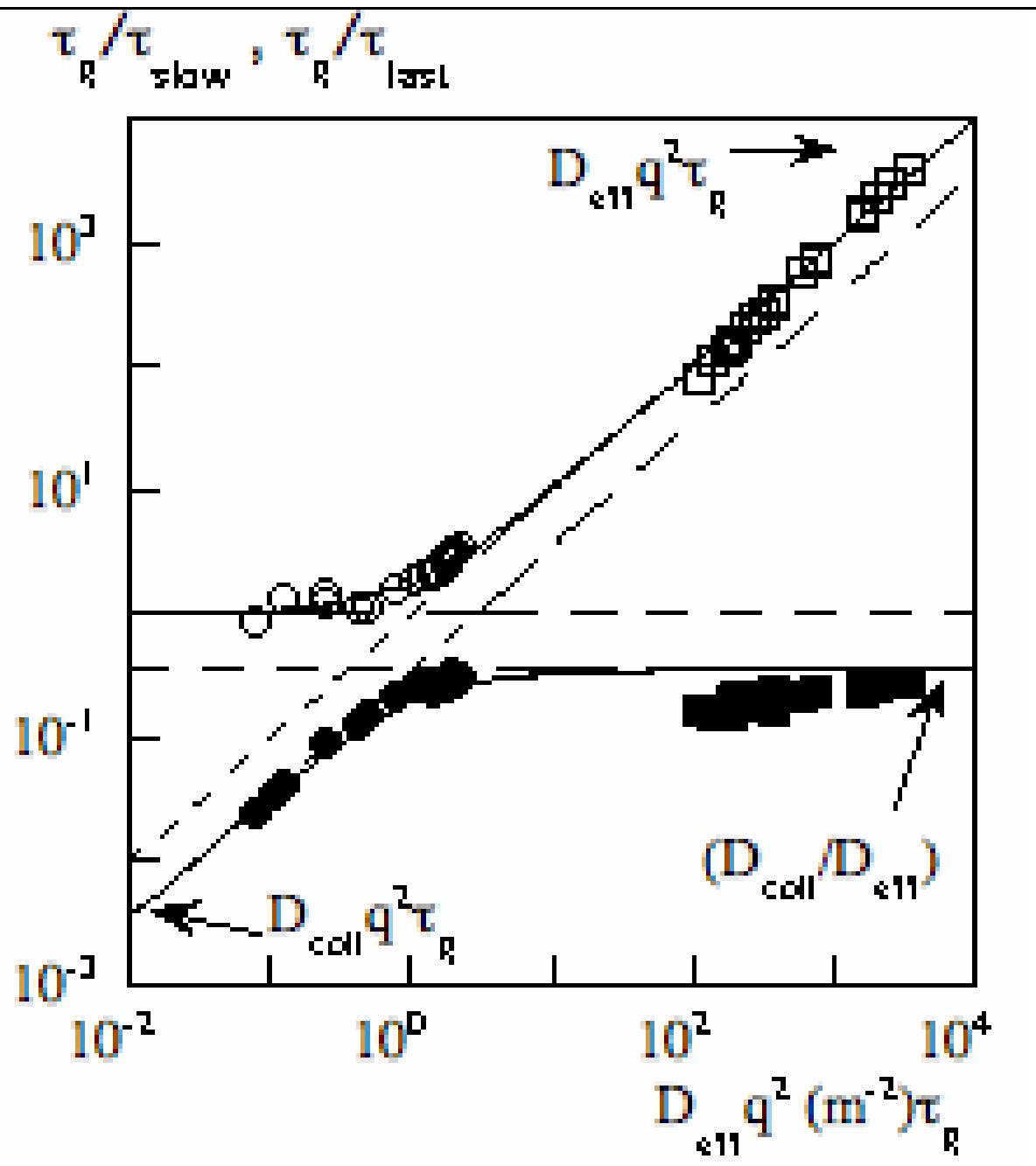

Figure 8 Normalized inverse characteristic times as a function of $q^2$ in reduced units. The lines are calculated from Eqs (3) with $D_{eff}$ =3.4 $10^{-11} m^2 s^{-1}$ and $D_{coll}$ = 1.2 $10^{-11}$ $m^2 s^{-1}$ . The experimental data are circles for the C12-sample ($\tau_R$=8.1 $10^{-5}$ s) and squares for the C18-sample ($\tau_R$ =0.125s). Full (open) symbols correspond to the slow (fast) mode.

The relative amplitudes of the two modes in each sample are compared in fig. 9 to the curves calculated from Eqs (3) ; the experimental relative amplitudes for the C12-sample have been estimated regardless of the amplitude of the third mode which accounts for about 10% of the overall intensity and is constant over the explored q-range. The relative amplitudes are plotted on a linear scale and the agreement with the theory is reasonable. Quantitatively the agreement is good in the two limits of small and large $D_{eff} q^2 \tau_R$ but, in the intermediate regime, the two sets of experimental data don't merge together. In fact the model calculation underestimates the contribution of the slow mode for the C12-sample and overestimates it for the C18-sample.

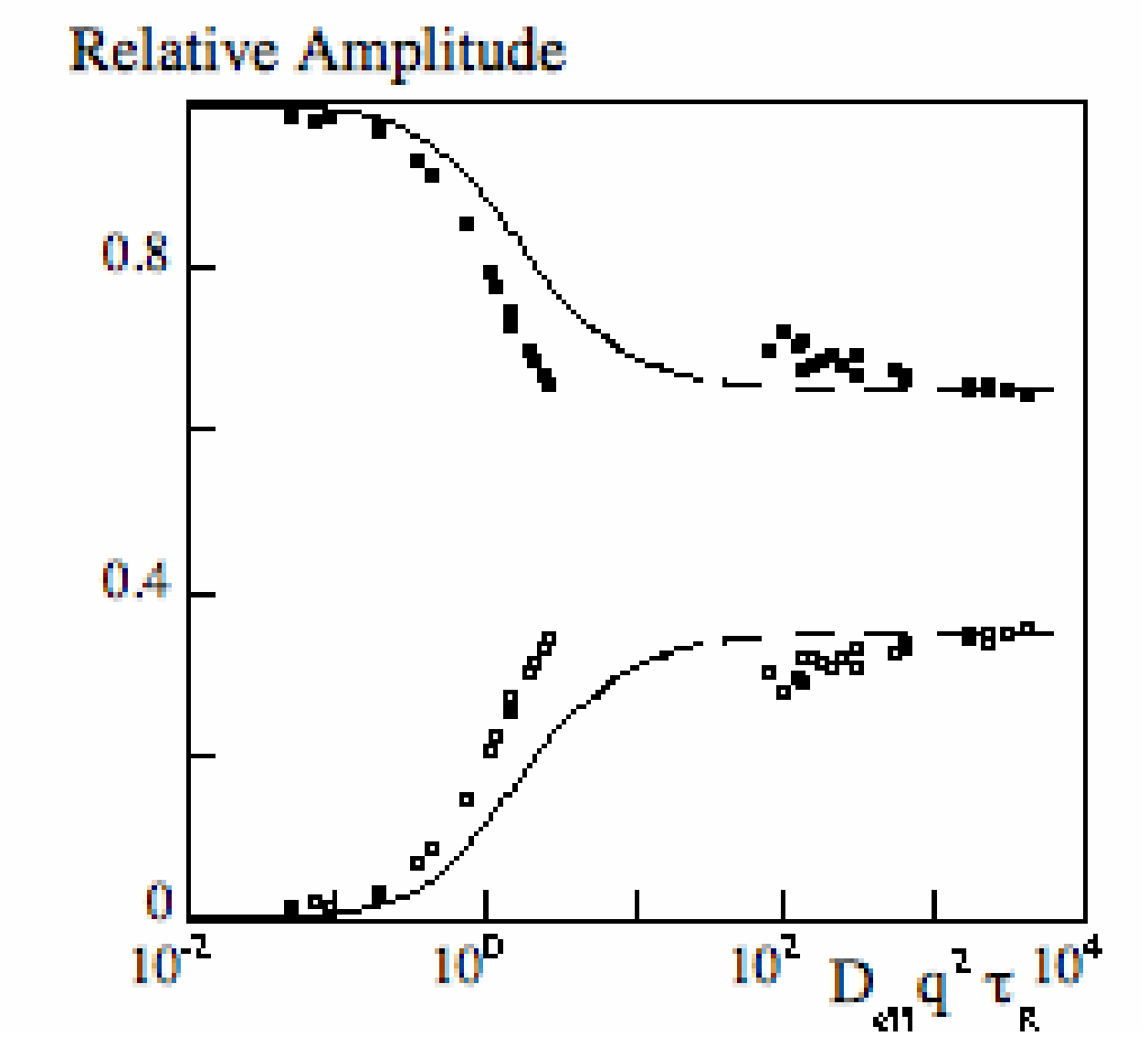

Figure 9 Relative amplitudes $\alpha_{f/s} / (\alpha_f + \alpha_s)$ as a function of $q^2$ in reduced units. The lines are calculated by Eqs(3) with $D_{eff}$=3.4 $10^{-11} m^2 s^{-1}$ and $D_{coll}$ = 1.2 $10^{-11}$ $m^2 s^{-1}$. The experimental points are circles for the C12-sample ($\tau_R$=8.1 $10^{-5}$ s) and squares for the C18-sample ($\tau_R$ =0.125s). Full (open) symbols correspond to the slow (fast) mode.

These slight discrepancies between theory and experiment remain to be explained. A possible explanation could be perturbations brought about by the third relaxation mode which is expected, as argued in the introduction, in view of the fact that a proper description of the system would imply three and not only two parameters. Another explanation could be that in DLS the viscoelastic properties are measured on a scale ( 300 to 3000 Å) not so large compared to the characteristic scale of the network (radius of the drops = 84Å and center to center distance of the drops 240Å) so that the coarse graining description of the viscoelastic properties must be refined by some kind of renormalization (i.e. for the elastic modulus or the terminal time measured macroscopically). This is an open question.

In the two fluids-model the parameters used to describe the viscoelastic properties of the system are the terminal time $\tau_R$ and the elastic modulus $G_O$ as measured in a rheological experiment. A final test is thus to examine how the parameters obtained here compare to the rheological values.

As regard the $\tau_R$'s, the injection of the rheological value for the C18-sample led us to a very consistent picture while the value for the C12-sample derived from an adjustement of the DLS data is in the predicted ratio to that of the C18-sample. Its low value is in line with the low viscosity of the C12-sample .

In the frame of the model outlined above, the elastic modulus $G_o = D_{el} n_o f/2$. From the experimental values given in table 2 or from the amplitude ratio we can deduce $\boldsymbol{D_{el}}$ = 2.2 ± 0.2 $10^{-11} m^2 s^{-1}$. The mean number density of drops, $\boldsymbol{n_o}$ is readily estimated from the size of the drops and the volume fraction .The frictional coefficient $\boldsymbol{f}$ is given by Stokes'law (see above) in the limit of high dilution but is modified by the hydrodynamic interactions at higher volume fractions . To take this into account we note that the collective diffusion coefficient can be written as $D_{coll} = \phi v \chi_T^{-1} / f$ with $\chi_T$ the osmotic compressibility: and $v$ the volume of the particules. The osmotic compressibility is deduced from the small angle neutron scattering results on the samples following the method described in ref 32 $\chi_T$ = 2 ± 0.2 $10^{-5}$ $Pa^{-1}$. This together with the measured value of $\boldsymbol{D_{coll}}$ allows for an independent estimate of the frictional coefficient $\boldsymbol{f}$ = 1.8 ± 0.3 $10^{-9}$ kg .$s^{-1}$. So that we eventually obtain $\boldsymbol{G_O}$ = 1400 ± 400 Pa in good agreement, within experimental errors, with the rheological value (1830 ± 100 Pa).

In conclusion, the two fluids-model, extended to Maxwellian fluids, describes quantitatively and coherently the dynamic properties of transient networks as probed by DLS. Furthermore a quantitative agreement is found between the values of the viscoelastic parameters $\tau_R$ and $\boldsymbol{G_O}$ obtained here and by rheological measurements. Our study illustrates the intricate dynamical behavior of such transient networks as probed by light scattering. In particular, the results for the C12 sample illustrate the intimate mixing of the collective diffusion and of the viscoelastic character that can be observed in the intermediate regime between the hydrodynamic and the gel limits . It is worth noting that DLS is an alternative method for the determination of $\tau_R$ and $\boldsymbol{G_O}$ , especially when $\tau_R$ is so small that conventional rheological measurements are difficult to achieve.

ACKNOWLEDGMENT:

We thank Raymond Aznar for synthetizing the telechelic polymer used in this work.